\documentclass[aps,prl,twocolumn,showpacs,floatfix]{revtex4}
\usepackage{graphics}
\usepackage{epsfig}
\usepackage{subfigure}
\usepackage{amsmath,amsfonts,amssymb,graphicx}

\begin{document}
\title{Molecular spring: from spider silk to silkworm silk}
\author{Xiang Wu$^{1,2}$, Xiang-Yang Liu$^{1,3\dagger *}$, Ning Du$^1$, Gang-Qin Xu$^1$ \& Bao-Wen Li$^{1,2,3\dagger}$}\email{phyliuxy@nus.edu.sg, phylibw@nus.edu.sg}
\affiliation{$^1$Department of Physics, Faculty of Science, National
University of Singapore, Singapore, 117542\\
$^2$Centre for Computational Science and Engineering, National University of Singapore, Singapore, 117542\\
$^3$NUS Graduate School for Integrative Sciences and Engineering,
Singapore, 117597, Republic of Singapore}
\date{\today}

\begin{abstract}
In this letter, we adopt a new approach combining theoretical
modeling with silk stretching measurements to explore the mystery of
the structures between silkworm and spider silks, leading to the
differences in mechanical response against stretching. Hereby the
typical stress-strain profiles are reproduced by implementing the
newly discovered and verified ``$\beta$-sheet splitting'' mechanism,
which primarily varies the secondary structure of protein
macromolecules; Our modeling and simulation results show good
accordance with the experimental measurements. Hence, it can be
concluded that the post-yielding mechanical behaviors of both kinds
of silks are resulted from the splitting of crystallines while the
high extensibility of spider dragline is attributed to the tiny
$\beta$-sheets solely existed in spider silk fibrils. This research
reveals for the first time the structural factors leading to the
significant difference between spider and silkworm silks in
mechanical response to the stretching force. Additionally, the
combination of theoretical modeling with experiments opens up a
completely new approach in resolving conformation of various
biomacromolecules.
\end{abstract}
 \maketitle

Natural silk spinning was pioneer of economy flourishing worldwide
since the first industrial revolution, and was continuing acting as
a pillar industry during the last several centuries until massive
manufacturing of man-made fibres derived from feedstocks of
petrochemical was popularized. Spider silk, superior to other
biomaterials, has extraordinary strength comparable to steel, and
the highest toughness among all natural silk fibres up to
date\cite{toughness1, toughness2, toughness3, toughness4}. It is
suggested that one strand of pencil thick spider silk can stop a
Boeing $747$ in flight. The exceptional properties serve new
application in industry to fulfill various functions, such as
bullet-proof vests, reinforced composites and aircrafts panels
substitute. On the other hand, the high production of silkworm silk
makes its irreplaceable stay in textile and other markets, despite
its inferior properties to spider silk\cite{strength, comparison2}.
Complementarily, the sustainable application has been penetrated in
daily life, ranging from costume manufacture to clinical treatment.
Both two kinds of silks exhibit greater environmental friendliness
and bio-compatibility than man-made petrochemical materials, which
implies the convenience of its synthesis, fabrication and
recycle\cite{compatible1, compatible2}. Although spider and silkworm
silk share a high degree of similarity in their chemical composition
and microscopic structures, the two kinds of silks behave
differently in their mechanical responses\cite{hierarchy1, beta1,
beta2, beta3}. To fulfill their potential functionalities requires a
full understanding of the underlying mechanism; however, to date
very few works have been engaged in this exploration.

Both spider silk and silkworm silk are primarily comprised of
Alanine- and Glycine-rich polypeptides in the fibrils; Besides,
natural spun silkworm silk contains two strands coated with sericin,
and the spider dragline silk contains major ampullate(MA) and minor
ampullate(MI) silk. X-ray diffraction and AFM probing\cite{spider1}
provide direct evidence on the molecular level that subtle deviation
in amino acids assembly sequence may lead to significant difference
in mechanical properties\cite{molecular1,
molecular2, molecular3}. 
However, it is still not sufficient to fully explain the mechanical
difference. As can be seen, the stress-strain profile of silkworm
silk is segmented in two regions, separated by yielding point. In
low stress regime, the silk fibre behaves linearly
elastic\cite{rubber1}, and skips into the post-yield region with
glass state which is nonlinear and irrecoverable\cite{molecular4}.
Different from silkworm silk, spider draglines show the
work-hardening phenomenon in the post-yield region, describing its
dramatic increase of the elasticity of spider silk when subjected to
certain extent of stretch, after which the fibres turn softer again.

Accordingly, the mechanical differences of spider draglines and
silkworm silk are also related to the high-order structures of
protein macromecules\cite{comparison1, strength, comparison2,
molecular2, secondary1}. The key problem is to determine structural
factors, which affect mechanical response to the external stress and
cause the differences of the two kinds of silks. The secondary
structure of silks can be divided into crystalline, primarily the
$\beta$-crystallites, and non-crystalline (amorphous) domains
(Fig.1(a), (b)), which contains random coils, $\alpha$-helices and
$\beta$-sheets\cite{beta, alpha}. It is believed that the secondary
structures are similar in both two kinds of silk fibres except that
ordered $\beta$-sheets were solely observed only in the amorphous
region of spider draglines\cite{spider1, secondary1, secondary2};
certainly, the crystallinity, denoting fraction of crystalline
region, and the sizes of $\beta$-crystallites are different. We
notice that one $\beta$-sheet in the amorphous region of spider
draglines are stacked by part of a single protein macromolecule,
while the $\beta$-crystallites in the crystalline regions are formed
by several adjacent of them. In order to distinguish the two kinds
of $\beta$-sheet structures, the one in crystalline region is
referred to the ``\textit{inter-protein $\beta$-crystallite}'', and
the one in amorphous region is referred to the
``\textit{intra-protein $\beta$-sheet}'', as shown in Fig.1(a) and
(b).

In this letter, aiming at revealing the connection between the
mechanical properties of spider and silkworm silk and their nano
structures, we are inclined to understand the roles of the elements
based on our models. Therefore, we are observing the collective
behaviors of two kinds of silk, in terms of the structural
difference, typically for the stress-strain profile.

\begin{figure}
\scalebox{0.45}[0.45]{\includegraphics{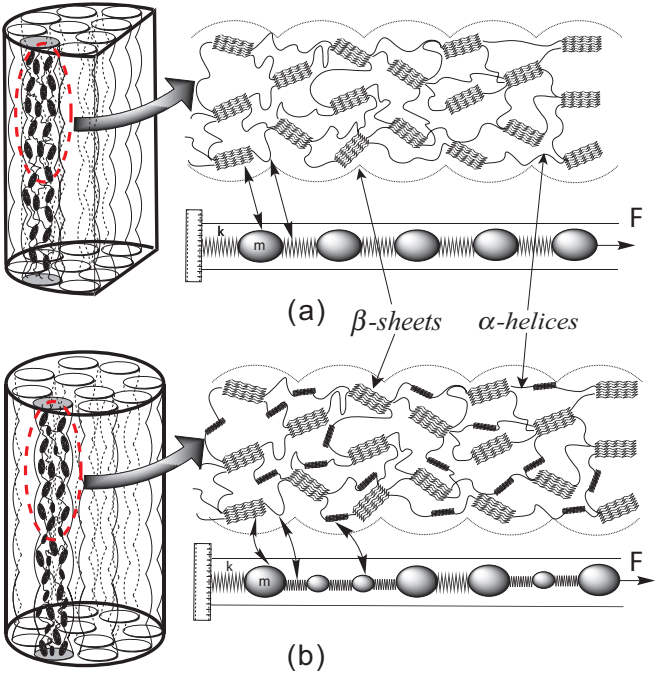}}
\caption{\label{fig:epsart}The left panels are the structures of
fibrils aligning along silkworm silk and spider silk fibres in (a)
and (b) respectively. The upper-right sketches are the
nano-structures of each fibril; the larger blocks denote the
inter-protein $\beta$-crystallites, and the smaller beads denote the
intra-protein $\beta$-sheets; the flexual lines in-between denote
helical or random structures in the amorphous matrix. The
lower-right parts is the model construction based on their
nano-structures, and the element correspondence is directed by the
thin solid arrows.}
\end{figure}

Let us first start our modeling of silkworm silk. As shown in
Fig.1(a), $\beta$-crystallites are extracted as massive bodies, and
the in-between springs represent the amorphous matrix, including
random coils, $\alpha$-helices, \textit{etc}. A quasi-periodic
structure(Figure 1a) is adopted with repeated segments of rigid
bodies and springs, and they are aligned along fibre axis subjected
to the one dimensional confinement we simplified to. As indicated by
Raman spectrum experiments, the $\beta$-crystallites and amorphous
regions are more likely to connect in series\cite{beta2}, which is
well described by the quasi-periodic structure. In general, the
transition from rubber to glass state at the yielding point requires
well definition in both the two regions. Hence, Hooke's springs are
imposed to describe the linear elastic behavior contributed mostly
by amorphous region before reaching the yielding point, and become
elasto-plastic with nonlinearity and irreversibility in the
post-yield region. In addition, $\beta$-crystallites, previously
treated as rigid bodies, are now actually elastic with relatively
$4$ times high elasticity of the whole fibre\cite{beta1}. Taking
into account the facts of the transverse size of
$\beta$-crystallites shrinking and the irreversible process of
elasticity measurement for the pre-stretched silk fibres,
``$\beta$-sheet splitting''  turns to be possible mechanism in
characterizing the yielding behavior\cite{beta1, beta2}, and will be
discussed in the following context.

Based on aforementioned facts, the model can be explicitly described
as a serially connected system of $N$ segments, each of which is
comprised of massive body with mass $m$ and light spring with
original elasticity $k_0$. The segment at one end is stretched by an
applied force $F$ and the other boundary one is fixed immobile.
During the stretch process, for an arbitrary segment $i$, the
extension linearly increases with the responsive force $F(i)$ in the
linear region until $F(i)$ reaches a criteria of threshold force
$F_{th}(i)$ , after which the $\beta$-crystallites start to split.
The collective behavior of splitting at $\{F_{th}(i)\}$
characterizes the critical behavior around the yielding point.
Because the splitting of $\beta$-crystallites requires external
energy input equal to or greater than the cohesive energy of
hydrogen bonds, equivalently a large applied force $F$ is likely to
break more hydrogen bonds, thus to retrieve longer amorphous protein
molecules from the compactly stacked $\beta$-crystallites. Hereby, a
simple linear relation characterizing such observation between
applied force $F$ and the released molecule length $\Delta L(i)$ can
be presumed as:

\begin{equation}
\Delta L(i)=\begin{cases}0 & F<F_{th}(i)\\
(F-F_{th}(i))/E_0 & F\geq F_{th}(i)
\end{cases}
\end{equation}

Therefore, the relation between applied force $F$ and observable
extension $\Delta x$ ($=\sum_{i=1}^{N}\Delta x_i$) can be deduced in
Eq.(2) as:

\begin{eqnarray}
 F&=& k_{\text{eff}}\cdot (\Delta x-\Delta L)\\
\nonumber &=& \frac{1}{\sum_{i=1}^{N}\frac{1}{k_i}}\cdot(\Delta
x-\sum_{i=1}^{N}\Delta L(i))\\
  &=&
\nonumber  \frac{k_0E_0L_0\Delta
x-Nk_0L_0\cdot\int_{0}^{F}(F-f)\rho_{th}(f)\texttt{d}f}{NE_0L_0+N\int_{0}^{F}(F-f)\rho_{th}(f)\texttt{d}f},
\end{eqnarray}

\noindent where $\rho_{th}(\cdot)$ is the distribution of the
threshold force $\{F_{th}(i)\}$, $L_0$ is the original length of
each segment, and
$h(x)$($\int_{-\infty}^{x}{\delta(x)\tt{d}\textit{x}}$) is the step
function. Besides, the calculation of elasticity $k_{\text{eff}}$
follows the law of sequentially linked springs with effective
elasticity of each segment $k_i=k_0/(1+\Delta L(i)/L_0)$.
Additionally, for large system size $N$, continuity approximation is
applied in deriving Eq.2. Mapping three dimensional reality to our
one dimensional model, the scaling is governed by such regulations:
$F\sim\sigma$(set the area of cross section $\Delta S=1$), and
$\Delta x\sim\epsilon\cdot NL_0$, where $\sigma$ and $\epsilon$
denote the tensile stress and tensile strain respectively. As one of
the key features of biomaterial, ununiformity caused by random
generated defects leads to serious instability of their mechanical
properties. Without loss of generality, gaussian random function is
chosen for the distribution of threshold forces.

\begin{figure}
\centering
\scalebox{0.80}[0.80]{\includegraphics{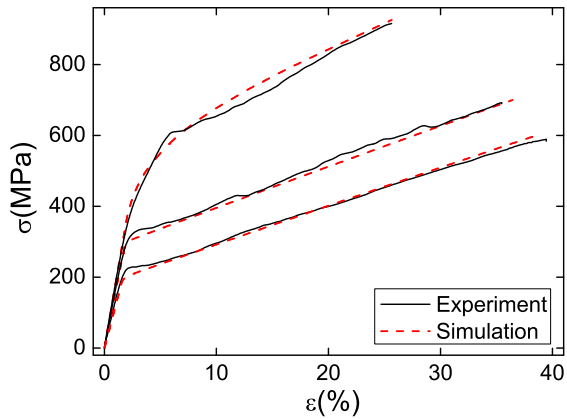}}
\caption{\label{fig:epsart} The relation between stress $\sigma$ and
strain $\epsilon$ of silkworm silk fibres: the black solid line is
the experimental curve, and the red dashed line is the simulation
results. A computer-controlled motorized spindle was used to draw
fibers from silkworm with $12mm/sec$, $24mm/sec$ and $36mm/sec$ from
the bottom up. Two successive $45min$, $95^\circ C$ heating
procedures in $0.5\%$ $Na_2CO_3$ and $1\%$ soap solution was adopted
in the degumming process. Measurements were performed using an
Instron MicroTester (Model 5848; force resolution, $0.5\%$ of
indicated load; position resolution $0.02\mu m$; strain rate is
$50\%/min$), at $20\pm 2^\circ C$ and the humidity was kept at
$60\pm 5\%$. the scaling constant in simulation $S_c=20$, the
corresponding parameters $k_0$, $E_0$, and $\langle F_{th}\rangle$
increase with the reeling speed.}
\end{figure}

\begin{figure}
\centering
\includegraphics[angle=0,scale=0.9]{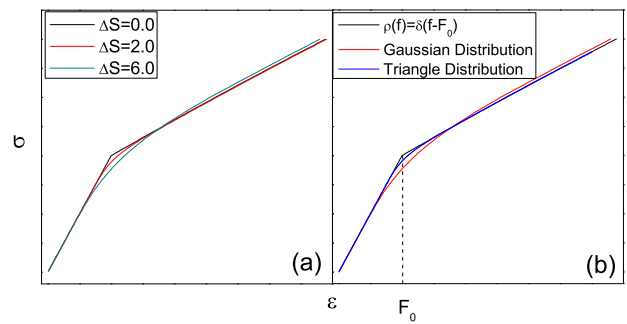}
\caption{\label{fig:epsart} (a):Stress-strain profile of silkworm
silk for different spread $\Delta S$ of the gaussian form
$\rho_{th}(\cdot)$ and same expectation value $\langle
F_{th}\rangle$. (b):Stress-strain profile of silkworm silk for
different forms of $\rho_{th}(\cdot)$ with same $\langle
F_{th}\rangle$ and various spread.}
\end{figure}

To mimic the dynamic process of silk stretching, the extensive
numerical simulations based on Molecular Dynamics (MD) were carried
out. In comparison, the stress-strain profiles of silkworm
silk(Fig.2) and spider draglines(Fig.4) under different reeling
speeds were measured. As shown in Fig.2, MD numerical results of
stress-strain profiles coincide very well with experimental data at
the proper values $E_0/k_0$ and $\langle F_{th}\rangle$. It is found
that the ratio $E_0/k_0$ determines the relative steepness of the
two-segmented stress-strain profiles, with two limits that the
post-yield behavior is strain-independent for $E_0/k_0\sim 0$, and
remains linear elastic for $E_0/k_0\sim \infty$. The stress-strain
profile is relatively robust to the form of distribution function of
threshold forces, as general gradually varying and centralized
functions give out similar profiles. As shown in Fig.3, various
distribution of threshold force $\rho_{th}$ gives no essential
difference, but only little variation. Additionally, different
reeling speed corresponds to their specific $\langle F_{th}\rangle$
and $E_0/k_0$, reflecting the initial formation difference of the
nano structures of macromolecules, presumably related to the
$\beta$-crystallites position and orientation. Therefore, the
breaking of $\beta$-crystallites in silkworm silk fibrils weakens
the linkage between protein molecules, which leads to its softening
in the post-yield regime.

To model the structure of spider dragline, we take into account the
performance of the intra-protein $\beta$-sheets in the amorphous
region (Fig.1(b)), which is uniquely observed spider draglines, in
addition to adopting the idea of the above model. Analogous to the
silkworm silk modeling, the linear assumption of protein molecule
length split from intra-protein $\beta$-sheets $\{\Delta L'(i)\}$
versus stretching force $F$ is preserved, with proportional constant
$E_1$ and threshold forces of splitting $\{F'_{th}(i)\}$. Due to the
morphological inperfectness and stacking incompactness of the tiny
intra-protein $\beta$-sheets, they are easier to be split than the
inter-protein $\beta$-crystallites, thus to release longer protein
molecules given the same external energy consumption. Besides, the
limited size of intra-protein $\beta$-sheets lead to their complete
destruction during stretching at forces $\{F_{tr}(i)\}$. Hereby, if
the stretching force $F$ larger than this terminating force, the
intra-protein $\beta$-sheets can no longer release any length of
protein molecules; instead, they are fully destroyed, and behave
similarly as coiling structures in the amorphous matrix. Concerning
the above statements, qualitative relation can be derived as
$E_1<E_0$, and $\langle F'_{th}\rangle < \langle F_{tr}\rangle <
\langle F_{th}\rangle$, which should be well obeyed in the following
modeling. Force-extension relation of spider draglines can be
analogously derived as that of silkworm silk, by setting the similar
form of the lengthening term $\{\Delta L(i)\}$ for inter-protein
$\beta$-crystallites in Eq.1,  and a relatively more complicated
form for intra-protein $\beta$-sheets shown in Eq.3.

\begin{equation}
\Delta L(i)=\begin{cases}0 & F<F'_{th}(i)\\
(F-F'_{th}(i))/E_1 & F'_{th}(i)\leq F<F_{tr}(i)\\
(F_{tr}(i)-F'_{th}(i))/E_1 & F\geq F_{tr}(i)
\end{cases}
\end{equation}

The explicit force-extension expression of spider silk turns out to
be:

\begin{eqnarray}
F\approx\frac{k_0L_0\Delta
x-Nk_0L_0\cdot\Omega(F)}{NL_0+N\Omega(F)},
\end{eqnarray}

\noindent where $\Omega(F)$ denotes:

\begin{eqnarray}
\nonumber \Omega(F)&=&\frac{p}{E_1}\left(1-\int_{0}^{F}\rho_{tr}(f)\texttt{d}f\right)\int_{0}^{F}(F-f)\rho'_{th}(f)\texttt{d}f\\
\nonumber &+&p\cdot\int_{0}^{F}\int_{0}^{f_2}\left(\frac{f_2-f_1}{E_1}\right)\rho_{tr}(f_2)\rho'_{th}(f_1)\texttt{d}f_1\texttt{d}f_2\\
&+&\frac{1-p}{E_0}\int_{0}^{F}(F-f)\rho_{th}(f)\texttt{d}f.
\end{eqnarray}

Therein, $\rho'_{th}(\cdot)$, $\rho_{tr}(\cdot)$ and
$\rho_{th}(\cdot)$ respectively characterize the distribution of the
threshold forces, the critical forces terminating the intra-protein
$\beta$-sheets splitting, and the threshold forces for inter-protein
$\beta$-crystallites. Correspondingly, Eq.5 shows the contribution
from intra-protein $\beta$-sheets splitting, their complete
destruction and the splitting of inter-protein $\beta$-crystallites.
The occurrence of each mechanism is also reflected in the
segmentation of stress-strain profile for spider draglines.
Similarly as silkworm silk, simulations have been carried out to
reproduce the stretch process of spider draglines. As shown in
Fig.4, the parameter ratio $E_1/(pk_0)$ and $E_0/((1-p)k_0)$ are
dominant to characterize the relative steepness of stress-stain
profile. Besides, the stress-strain profile is robust to the exact
form of force distribution functions; instead, only their central
values denoting general each mechanism in effect need to be
determined. Moreover, it follows that linear elasticity $k_0$ and
proportional constant $E_0$ and $E_1$ increases with reeling speed,
indicating their dependence of initial conformation of structures.
Therefore, the mechanism of each segment of the stress-strain
profile of spider dragline has been clearly distinguished: the
intra-protein $\beta$-sheets split at the first yielding, and their
complete destruction gives rise to the lengthening of the fibrils
and the stiffening in the post-yield region, thus the work-hardening
phenomenon is well reproduced. On the other hand, the inter-protein
$\beta$-crystallites split during or after the release of
intra-protein $\beta$-sheets in the similar way of silkworm silk,
resulting in their softening before breaking. The diverse high-order
structures of protein molecules for silk fibres of different species
result into their distinct mechanical complexity.

\begin{figure}
\scalebox{0.80}[0.80]{\includegraphics{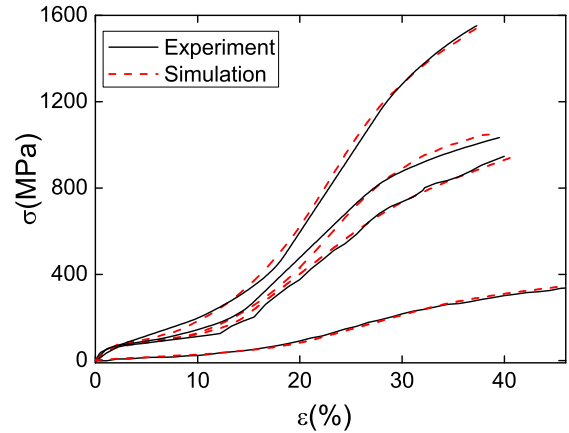}}
\caption{\label{fig:epsart} The black solid lines denote the
stress-strain profile for spider silk under different reeling
speed\cite{spider1}. Stress-strain curves of spider dragline silk
($2.5, 10, 25, 100mm/sec$ motor-reeled at $22^\circ C$, from the
bottom up) were performed using an Instron MicroTester (Model 5848;
force resolution, $0.5$\% of indicated load; position resolution
$0.02\mu m$; strain rate is $50\%/min$), at $22^\circ C$ and the
humidity was kept at $55-60\%$. The red dashed lines are the
computational results, where scaling constant $S_c=20$, the
preserved fraction of intra-molecule $\beta$-sheet $p=0.5$, and
parameters $k_0$, $E_0$ and $E_1$ are varied for different reeling
speeds.}
\end{figure}

In conclusion, we have established the correlation between the nano
structures and mechanical properties of silk fibres in our essay, on
the basis of modeling in combination with the silk stretching
experiments. We obtained for the first time the breaking mechanism
of the two types of silks. The splitting of inter-protein
$\beta$-crystallites gives rise to the weakening the linkage among
molecules while limited capacity of intra-protein $\beta$-sheets
results in the lengthening and the extra stiffening in the
post-yield regime of spider draglines. Moreover, the exploration
from molecular mechanical properties turns out be a brand new
probing method to investigate the subtle structural differences
among biomacromolecules. This provides a new passage in
understanding the origin of molecular mechanical behaviours, and
will facilitate the identification of robust technologies in
fabricating silks of ultrafunctionality.

\end{document}